\documentclass[english]{article}

\usepackage[T1]{fontenc}
\usepackage[latin9]{inputenc}

\usepackage{cite}
\usepackage{amsmath}
\usepackage{graphicx}

\usepackage[unicode=true]{hyperref}

\usepackage{setspace}
\usepackage[a4paper]{geometry}
\makeatletter

\let\oldbibliography\thebibliography
\renewcommand{\thebibliography}[1]{%
  \small%
  \oldbibliography{#1}%
  \setlength{\itemsep}{0pt}%
}

\makeatother

\begin{document}
\title{Noisy voter model for the anomalous diffusion of parliamentary presence}
\author{A. Kononovicius\thanks{email: \protect\href{mailto:aleksejus.kononovicius@tfai.vu.lt}{aleksejus.kononovicius@tfai.vu.lt};
website: \protect\url{http://kononovicius.lt}}}
\date{Institute of Theoretical Physics and Astronomy, Vilnius University}
\maketitle
\begin{abstract}
We examine parliamentary presence data of the 2008--2012 and the
2012--2016 legislatures of Lithuanian parliament. We consider cumulative
presence series of each individual representative in the data set.
These series exhibit superdiffusive behavior. We propose a modified
noisy voter model as a model for the parliamentary presence. We provide
detailed analysis of anomalous diffusion of the individual agent trajectories
and show that the modified model is able to reproduce empirical statistical
properties.
\end{abstract}

\section{Introduction}

Numerous processes observed in variety of physical systems have been
known to diffusive faster or slower than the classical Brownian particle.
This family of process is often referred to as anomalous diffusion.
In one dimensional case the anomalous diffusion would be characterized
by the root mean square displacement (standard deviation) of the following
dependence on time:
\begin{equation}
\sqrt{\left\langle \left(\Delta x\right)^{2}\right\rangle }\sim t^{\alpha},
\end{equation}
where $\Delta x$ is the distance the diffusing particle has moved
away from the origin with the average being taken over ensemble of
particles. For the classical Brownian motion (normal diffusion) we
would have $\alpha=\frac{1}{2}$. A process with slower diffusion,
$\alpha<\frac{1}{2}$, would be considered to be subdiffusive. Subdiffusion
is often assumed to be caused by the particles jumping from one local
minima, being trapped for a prolonged period of time and then jumping
to another local minima \cite{Metzler2000PhysRep,Chepizko2013PRL,Evers2013PRE,Iaconis2019PRB}.
Processes exhibit superdiffusion if the diffusion is faster than the
normal diffusion, with $\alpha>\frac{1}{2}$. Superdiffusion is often
assumed to be observed in diffusive processes exhibiting Levy flights
\cite{Metzler2000PhysRep,Mercadier2009NatPhys,Scalliet2015PRL,Kiselev2019PRL}.
Another possible causes behind anomalous diffusion could be time subordination
\cite{Fogedby1994PRE,Baule2005PRE,Kazakevicius2015PhysA,Ruseckas2016JStat}
and heterogeneity in the media \cite{Cherstvy2013NJP,Cherstvy2014SM,Kazakevicius2016PRE}.
Here we will consider anomalous diffusion in the parliamentary presence
data. Similar analysis was already conducted by \cite{Vieira2019PRE}
using Brazilian parliamentary presence data. Reportedly, strong evidence
for the ballistic regime, $\alpha\approx1$, was found. In regards
to anomalous diffusion our approach to analysis is mostly similar,
but we use Lithuanian parliamentary presence data. Furthermore we
also consider other statistical properties of the empirical data,
such as attendance streak distributions, which provide additional
information about the process as well as additional way to validate
a model.

In \cite{Vieira2019PRE} a phenomenological model for the parliamentary
presence was proposed by the means of a non--linear diffusion equation.
Here we propose an agent--based model for the parliamentary presence.
At its core the proposed agent--based model is the voter model. The
voter model and variety of its modifications have been under active
consideration by opinion dynamics (sociophysics) community \cite{Castellano2009RevModPhys,Jedrzejewski2019CRP,Redner2019CRP}.
Such as the impact of inflexibility \cite{Mobilia2007JStatMech,Khalil2018PRE},
spontaneous flipping \cite{Kirman1993QJE,Granovsky1995}, variety
of network topology effects \cite{Alfarano2009Dyncon,Kononovicius2014EPJB,Carro2016,Peralta2018Chaos,Mori2019PRE,Gastner2019JPA},
private opinions \cite{Masuda2010PRE,Gastner2018JStat,Jedrzejewski2018PlosOne},
nonlinear interactions \cite{Artime2019CRPhys,Castellano2009PRE}
were studied in the framework of the voter model. Various voter models
were applied to model electoral and census data \cite{FernandezGarcia2014PRL,Sano2016,Kononovicius2017Complexity,Braha2017PlosOne,Kononovicius2019CompJStat}
as well as to model financial markets \cite{Alfarano2005CompEco,Alfarano2008Dyncon,Kononovicius2012PhysA,Gontis2014PlosOne,Franke2018,Kononovicius2019OB,Vilela2019PhysA}.
Anomalous diffusion, to the best of our knowledge, was not studied
in any of the voter models, because these models should not exhibit
anomalous diffusion. Yet our approach takes a different point of view
than is common in the analysis of the voter models, here we consider
individual agent trajectories. From this point of view observing anomalous
diffusion is quite possible.

Our goal in this paper is to understand anomalous diffusion in the
parliamentary presence data in the context of the voter models. Having
this goal in mind we have organized the paper as follows. In Section~\ref{sec:empirical-analysis}
we conduct empirical analysis, which helps us to provide context for
the numerical modeling of the parliamentary presence phenomenon. In
Section~\ref{sec:model} we describe the noisy voter model and its
modifications. In Section~\ref{sec:analysis} we analyze anomalous
diffusion of the individual agents trajectories and show that the
model is able to reproduce the empirical observations. We provide
concluding remarks and discussion in Section~\ref{sec:conclusions}.

\section{Empirical analysis of the parliamentary presence data\label{sec:empirical-analysis}}

We have obtained the registration to vote data, which indicates willingness
of the representative to vote on the agenda, from the Lithuanian parliament's
website \cite{LRSData}. Based on the data we have constructed presence
time series, $\eta_{t}^{\left(i\right)}$, for each of the representatives
in the Lithuanian parliament (index $i$ loops through the representatives,
while index $t$ is the session number). We assume that representative
was absent during the session, if the representative did not register
to vote at all during that session, and encode this as $0$. Otherwise
we assume that representative was present and encode this as $1$.

The raw registration to vote data also indicates whether the representative
was elected (the seat taken) at the time of the session. Reasons why
a particular seat could be empty vary: death, prosecution or being
elected to a different post. While the replacements are elected as
soon as possible, we still have to deal with some missing data. Unlike
in \cite{Vieira2019PRE}, we detect the replacements and join the
respective presence time series. Suppose that representative A left
his seat after $t_{A}$ sessions, his possible replacements would
be all representatives who have taken their seats after $t_{A}$ sessions.
Among all representatives who were not present till the end of their
term, we find those with the least possible replacements (though the
number of possible replacements should be larger $1$). If there are
multiple possible replacements, we select the one who took their seat
the earliest and join the records of both representatives. We proceed
until all replacements are found (at this point the data set contains
$141$ records). This procedure should minimize the number of records
with the missing data, yet if at this point some data is still missing,
then we replace the records containing missing data with the copies
of valid records. Average replacement percentage for both considered
legislatures was around $10\%$. The reported results are robust in
respect to this random replacement procedure. Due to the replacement
scheme our data sets always have exactly $141$ presence time series
(as there $141$ seats in the Lithuanian parliament). We have made
the processed attendance data set for the legislatures of 2008--2012
and 2012--2016 available via GitHub repository \cite{Kononovicius2020GitPresence}.

\begin{figure}
\begin{centering}
\includegraphics[width=0.95\textwidth]{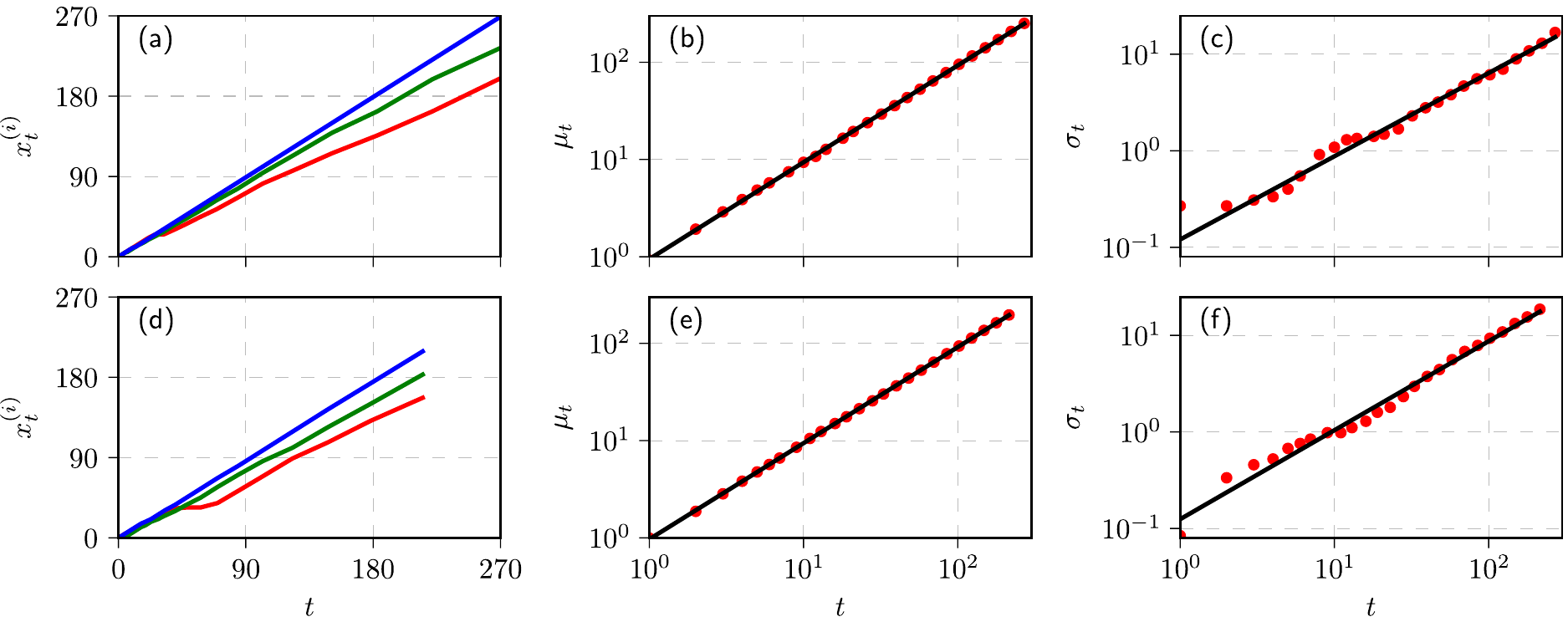}
\par\end{centering}
\caption{Exemplary presence time series ((a) and (d)), mean ((b) and (e)) and
standard deviation ((c) and (f)) series for the 2008--2012 legislature
((a), (b) and (c)) and 2012--2016 legislature ((d), (e) and (f))
data. Red dots ((b), (c), (e) and (f)) represent empirical data, while
black lines provide power--law fits with exponents $\alpha=1$ ((b)
and (e)) and $\alpha=0.85$ ((c) and (f)).\label{fig:empirical} }
\end{figure}

As in \cite{Vieira2019PRE} we take primary interest in the cumulative
presence series, which are obtained directly from $\eta_{t}^{\left(i\right)}$
series and are defined as:
\begin{equation}
x_{t}^{\left(i\right)}=x_{t-1}^{\left(i\right)}+\eta_{t}^{\left(i\right)}.
\end{equation}
Note that in the begging of each legislature we reset the attendance
record, $x_{0}^{\left(i\right)}=0$, for each representative. Using
the cumulative series we observe the temporal evolution of its mean
(over individual representatives at a particular time): 
\begin{equation}
\mu_{t}=\frac{1}{N}\sum_{i=1}^{N}x_{t}^{\left(i\right)},
\end{equation}
and its standard deviation (over individual representatives at a particular
time):
\begin{equation}
\sigma_{t}=\sqrt{\frac{1}{N-1}\sum_{i=1}^{N}\left[x_{t}^{\left(i\right)}-\mu_{t}\right]^{2}}.
\end{equation}
See Fig.~\ref{fig:empirical} for example of the empirical $x_{t}^{\left(i\right)}$
series and the empirical $\mu_{t}$ and $\sigma_{t}$ series. As one
should expect, due to the way we have encoded $\eta_{t}^{\left(i\right)}$
as well as due to high average presence rates ($\sim90\%$ in both
cases), the mean grows linearly with time $\mu_{t}\sim t$. While
the standard deviation clearly shows the superdiffusive behavior,
$\sigma_{t}\sim t^{\alpha}$ with $\alpha=0.85$. Note that the previous
analysis of the Brazilian parliament presence data \cite{Vieira2019PRE}
reported that the ballistic regime, $\alpha\approx1$, was found for
the standard deviation. This difference could be related to the different
treatment of the missing data as well as to the differences between
Lithuanian and Brazilian parliaments. There are other possible reasons
related to the individual level dynamics, which are discussed in the
following sections.

Closer examination of Fig.~\ref{fig:empirical}~(c) reveals that
some correlation in the residuals is present. This is likely related
to the limited amount of empirical data: both in temporal sense (our
data sets cover slightly more than $200$ parliamentary sessions)
and realization sense (our data sets include $141$ attendance records).

\begin{figure}
\begin{centering}
\includegraphics[width=0.7\textwidth]{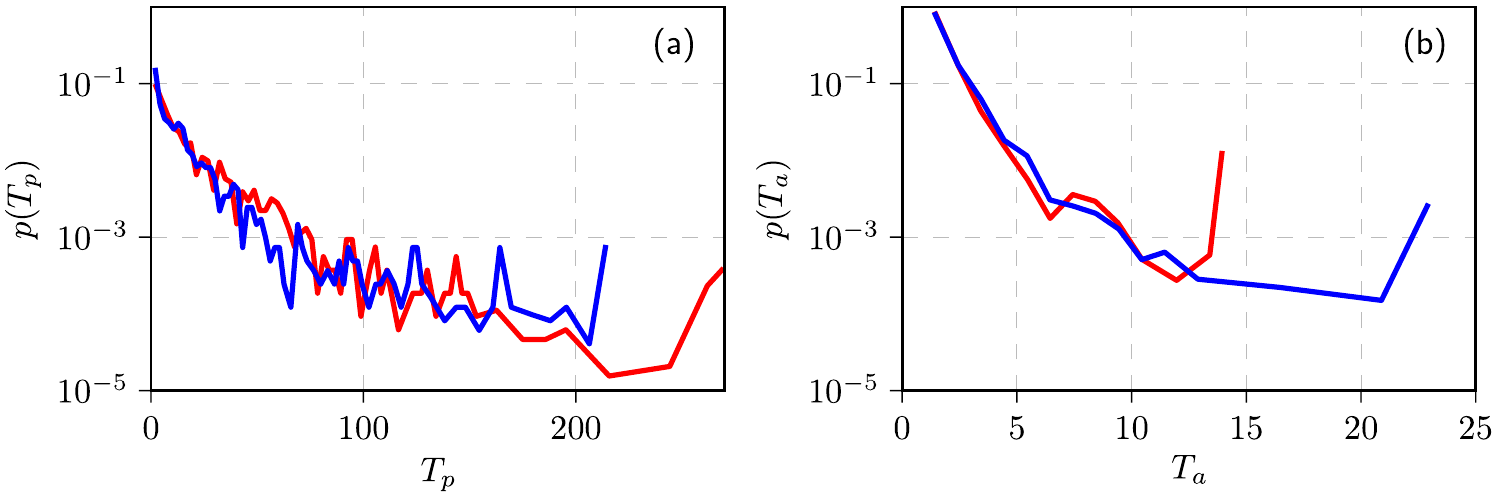}
\par\end{centering}
\caption{Probability density functions of the length of presence, $T_{p}$,
(a) and absence, $T_{a}$, (b) streaks in the presence data set for
the 2008--2012 (red curves) and 2012--2016 (blue curves) legislatures.\label{fig:empirical-bursts}}
\end{figure}

We supplement our empirical analysis by considering presence, $T_{p}$,
and absence, $T_{a}$, streaks of the individual representatives (see
Fig.~\ref{fig:empirical-bursts}). Shorter streaks, $T_{p}<100$
and $T_{a}<10$, seem to be distributed exponentially, suggesting
that underlying process could be a Poisson process. While the longer
streaks break the trend indicating that underlying process might be
a non--homogeneous Poisson process or a non--Poisson process. In
the next section we introduce an agent--based model driven by a Poisson
process, which is non--homogeneous in time.

\section{Noisy voter model of the parliamentary presence\label{sec:model}}

Original definition of a model, which is now known as the voter model,
involved only a simple replacement mechanism \cite{Clifford1973}.
The replacement mechanism was assumed to represent competition between
two species, but it can also represent competition between social
ideas or behaviors. In fact the model has found wider recognition
in opinion dynamics community and therefore is known as the voter
model \cite{Liggett1999,Castellano2009RevModPhys}. While it is quite
easy to imagine direct competition between two species, competition
between the ideas is indirect instead. It happens only because social
animals, e.g., humans, tend to exert social pressure on each other,
which is the force to causing the replacement of less popular ideas
by the more popular ones. Though, admittedly there are a few possible
ways to respond to social pressure \cite{Willis1965,Kirman1993QJE,Nail2016APPA},
which can have profound effects on the observed dynamics. We believe
that the voter model with noise is the simplest model, which directly
includes both social conformity and independence mechanisms and indirectly
takes into account anti--conformist behavior. In our earlier works
we have shown that the noisy voter model is quite applicable both
to finance \cite{Kononovicius2012PhysA,Gontis2014PlosOne,Kononovicius2019OB}
and opinion dynamics \cite{Kononovicius2017Complexity,Kononovicius2019CompJStat}.
Here we apply the noisy voter model to model the parliamentary presence.

Let us assume that after each session each member of the parliament
reconsiders his previous behavior. If the representative had intended
to attend (let us label this state as $1$), then the representative
could begin to intend to skip (let us label this state as $0$). Let
this transition occur with probability:
\begin{equation}
p_{1\rightarrow0}^{\left(i\right)}=h\left[\varepsilon_{0}+\frac{X_{0}}{N}\right].\label{eq:p10}
\end{equation}
Likewise, if the representative intended to skip, then the representative
could start to intend to attend. Let this transition occur with probability:
\begin{equation}
p_{0\rightarrow1}^{\left(i\right)}=h\left[\varepsilon_{1}+\frac{X_{1}}{N}\right]=h\left[\varepsilon_{1}+\left(1-\frac{X_{0}}{N}\right)\right].\label{eq:p01}
\end{equation}
In both of the transition probabilities above $h\cdot\varepsilon_{k}$
are the independent switching probabilities to the state labeled by
$k$, while $h\cdot\frac{X_{k}}{N}$ are the imitation switching probabilities
to the state labeled by $k$ (these transitions happen due to influence
of peers in the destination state). Effectively $h$ sets the rate
at which the agents change their state (the higher $h$ is the faster
the changes become), while $\varepsilon_{k}$ controls the impact
of peer pressure on the changes (the larger $\varepsilon_{k}$ the
more independent of peer pressure changes become). Due to conservation
of total number of agents, $N$, we have $X_{1}=N-X_{0}$ (here $X_{k}$
is the total number of agents in the state $k$). We consider only
those parameter values for which neither of the transition probabilities
for any $X_{0}\in\left[0,N\right]$ is larger than one.

Then just before the session each agent decides how to act (whether
to actually attend). Let the agent attend with probability $q_{k}$
given he is in the state $k$. In general $q_{k}$ can take any values
between $0$ and $1$, we only requite that $q_{1}\geq q_{0}$ as
agents in state $1$ are assumed to have an intent to attend.

\begin{figure}
\begin{centering}
\includegraphics[width=0.4\textwidth]{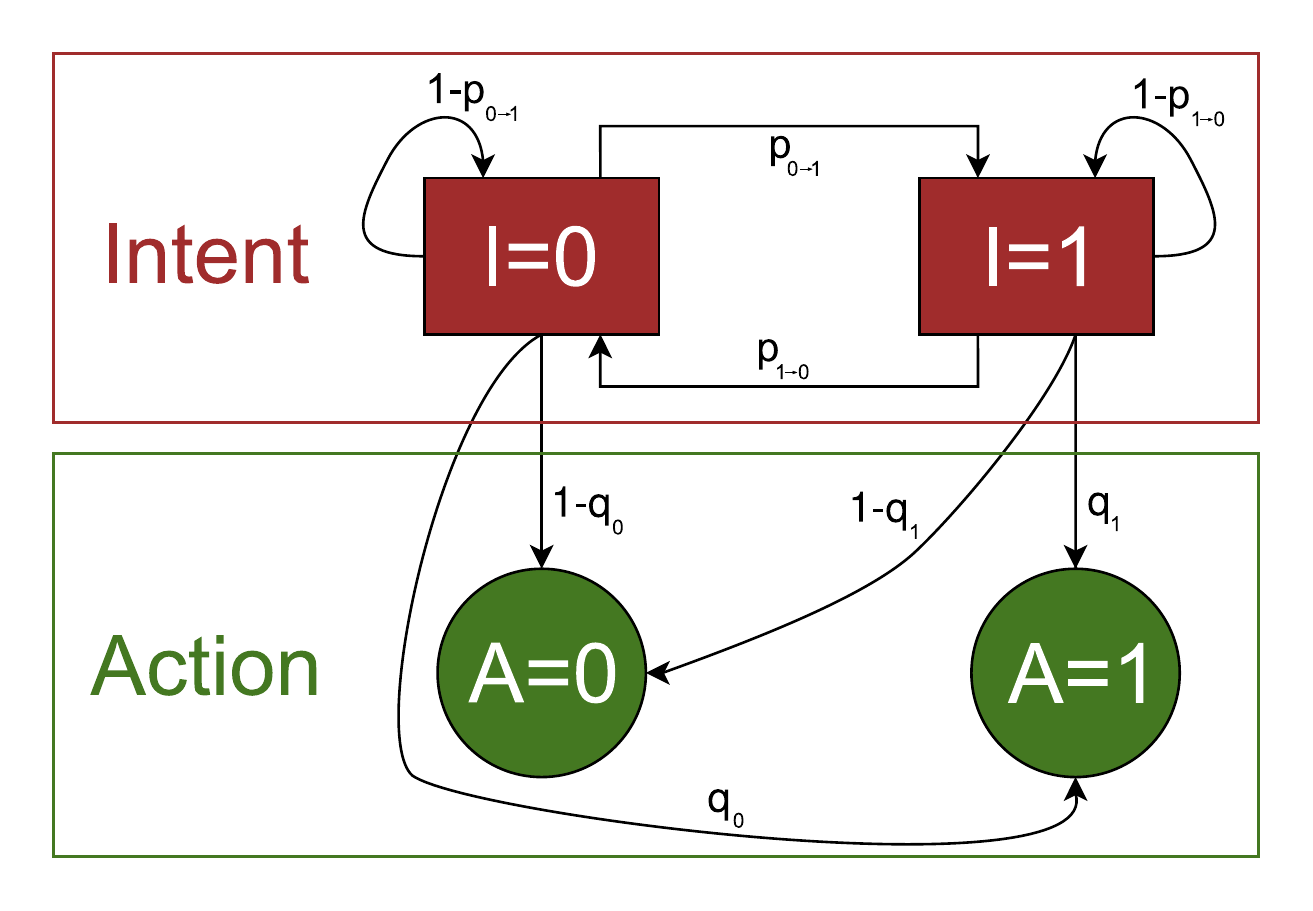}
\par\end{centering}
\caption{Representation of the model dynamics from an individual agent perspective.
Here $I$ represents the internal intent state and $A$ represents
the observed action state.\label{fig:hmm-logic}}
\end{figure}

This model can be seen as a special case of hidden Markov model \cite{Rabiner1989IEEE}.
Yet in our case each individual agent is described by its own hidden
Markov model: internal (intent) and observed (action) states describe
individual agents and not the whole system. In Fig.~\ref{fig:hmm-logic}
we have shown a representation of the model dynamics from an individual
agent perspective as hidden Markov model. It is important to note
that $p_{i\rightarrow j}$ depend on the intent of other agents, while
$q_{i}$ probabilities remain constant through out the simulation.

In this formulation of the noisy voter model more than one agent can
change its state after each time tick (which corresponds to a parliamentary
session in our case). This contrasts with the original formulation
of the voter model, which allows for just one agent to change state
during a single time tick. Original one--step formulation could be
seen as superior in a sense that it allows for the continuous time
treatment by the Gillespie method \cite{Gillespie1977JPC,Carro2015PlosOne}.
A similar formulation to the one used here was proposed in \cite{Goldenberg2001}
and compared against the Bass diffusion model, which is known to be
unidirectional (with one agent state being absorbing) variant of the
noisy voter model \cite{Daniunas2011ICCGI}. It was found that the
models produce very similar time series, but allowing for multiple
agents to switch per step introduces information lag into the model.

The model we have introduced here also bears certain similarities
to the voter models with private and public opinions \cite{Masuda2010PRE,Gastner2018JStat,Jedrzejewski2018PlosOne}.
Our approach is different, because we have the true state (or intent),
which is driven by the imitation behavior as it would be in any voter
model, and the observed state (or action), which is randomly taken
by the agent with probability depending on the true state. Another
similarity can be drawn to \cite{Martins2008IJMPC} in which agents
have continuous opinions, but act using discrete actions. In our approach
agents have discrete opinions as agent states are binary.

We have shared an implementation of the model in Python via GitHub
repository \cite{Kononovicius2020GitVM}.

\section{Anomalous diffusion of individual agent trajectories in the noisy
voter model\label{sec:analysis}}

Let us discuss the anomalous diffusion between individual agent trajectories
this model exhibits when applied as a model for parliamentary presence.
First of all for a variety of valid parameter sets we have observed
a linear trend in the mean series, $\mu_{t}$. While the trends of
the standard deviation series, $\sigma_{t}$, are a bit more sophisticated.

\begin{figure}
\begin{centering}
\includegraphics[width=0.7\textwidth]{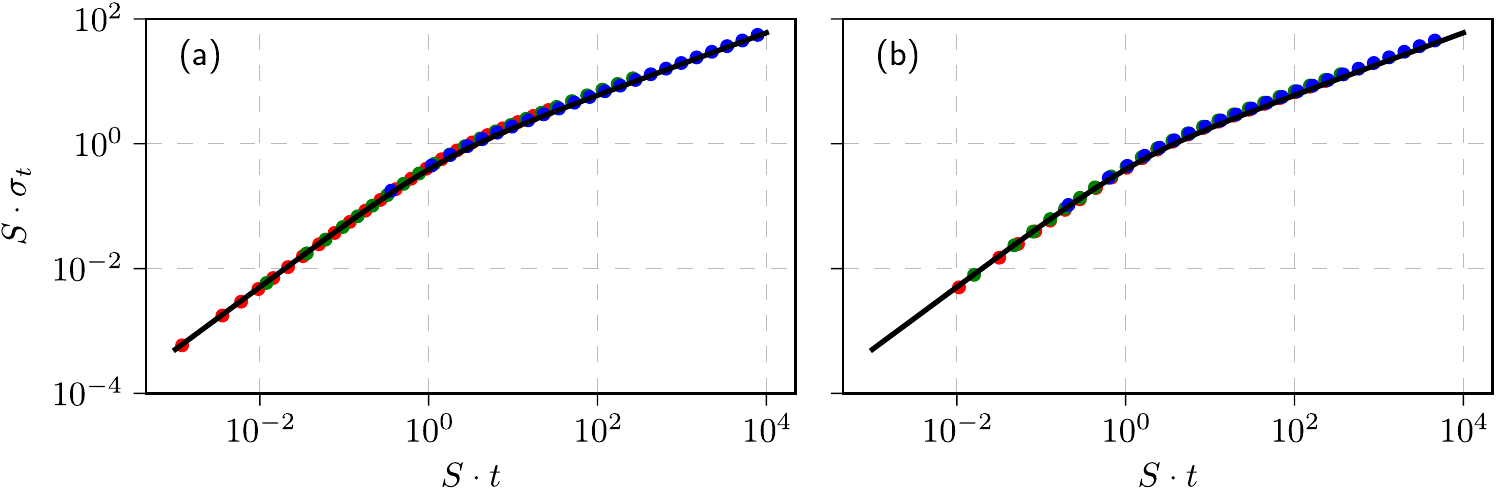}
\par\end{centering}
\caption{Scaling behavior of $\sigma_{t}$ in the simplest case (with $\varepsilon_{0}=\varepsilon_{1}=\varepsilon$,
$q_{0}=0$ and $q_{1}=1$) of the model: numerical results (colored
dots) and the scaling law, Eq.~(\ref{eq:stdFitLaw}), (black curves).
Scaling law parameter values: $\theta_{0}=0.66$ and $\theta_{1}=1.4$.
Default model parameter values: $\varepsilon=0.1$, $h=10^{-2}$,
$N=141$. Subfigure (a) shows the scaling behavior in respect to $h$
with the following values: $10^{-3}$ (red dots), $10^{-2}$ (green
dots) and $3\cdot10^{-1}$ (blue dots). Subfigure (b): $\varepsilon=0.03$
(red dots), $3$ (green dots) and $10$ (blue dots). \label{fig:stdSimpleScaling}}
\end{figure}

Let us start by considering the simplest case of the proposed model.
Namely, let us assume that the intent of agents is pure, i.e., they
either always skip, $q_{0}=0$, or attend, $q_{1}=1$, if they intend
to do so. Furthermore let us assume that the true states are equally
attractive for agents switching independently, i.e., let $\varepsilon_{0}=\varepsilon_{1}=\varepsilon$.
In this highly simplified case we observe that $\sigma_{t}$ exhibits
the following scaling behavior:
\begin{align}
\sigma_{t} & =\frac{\theta_{0}t}{\sqrt{\theta_{1}+St}}.\label{eq:stdFitLaw}
\end{align}
In the above $\theta_{0}$ and $\theta_{1}$ seem to be independent
of the model parameters (we estimate that $\theta_{0}=0.66\pm0.06$
and $\theta_{1}=1.4\pm0.55$), while $S$ seems to be fully determined
by the model parameters (the form was determined numerically),
\begin{equation}
S=h\left(1+2\varepsilon\right).
\end{equation}
Note that $S$ equals the sum of the transition probabilities. It
should be quite easy to see that on the shorter time scales the model
exhibits ballistic regime, $\sigma_{t}\sim t$, while on the longer
time scales normal diffusion, $\sigma_{t}\sim\sqrt{t}$, takes over.
In Fig.~\ref{fig:stdSimpleScaling} we have shown that this scaling
law rather nicely fits numerical results obtained with different values
of the model parameters. In Fig.~\ref{fig:stdSimpleScaling}~(b)
the difference between the numerical results representing two smallest
$\varepsilon$ is quite small. This is expected as $S$ changes very
little with $\varepsilon$ in those cases, because large $N$ dominates
the change in $\varepsilon$.

\begin{figure}
\begin{centering}
\includegraphics[width=0.4\textwidth]{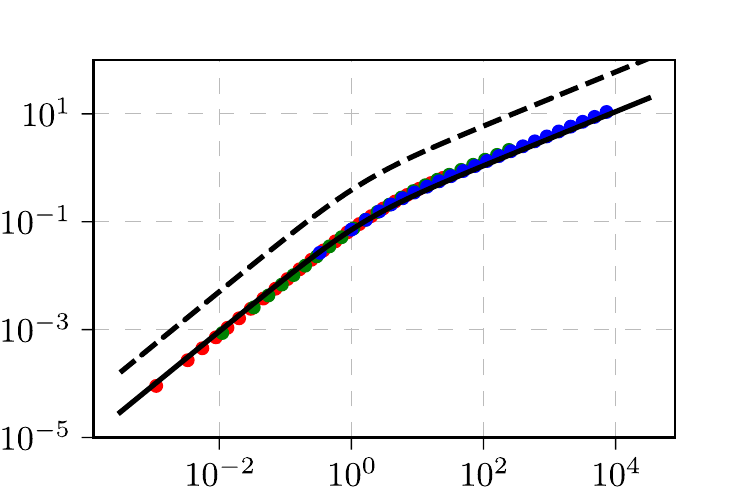}
\par\end{centering}
\caption{Scaling behavior of $\sigma_{t}$ in the asymmetric simplified case
(with $q_{0}=0$ and $q_{1}=1$) of the model: numerical results (colored
dots) and the scaling law, Eq.~(\ref{eq:stdFitLaw}), (black curves).
Scaling law parameter values: $\theta_{0}=0.66$ (dashed curve) and
$0.12$ (solid curve), $\theta_{1}=1.4$ (dashed curve and solid curve).
Model parameter values: $\varepsilon_{0}=0.1$, $\varepsilon_{1}=10$
and $N=141$ (all cases), $h=10^{-4}$ (red dots), $10^{-3}$ (green
dots) and $3\cdot10^{-2}$ (blue dots). \label{fig:stdAsymmScaling}}
\end{figure}

Breaking the symmetry assumption, i.e., allowing for $\varepsilon_{0}\neq\varepsilon_{1}$,
does not break the qualitative behavior of the scaling law. Though,
we need to rewrite the scaling multiplier as
\begin{equation}
S=h\left(1+\varepsilon_{0}+\varepsilon_{1}\right),
\end{equation}
and also $\theta_{0}$ value changes as the scaling law shifts downwards
(see Fig.~\ref{fig:stdAsymmScaling}). This downward shift is expected
as in the asymmetric case agents tend to prefer one state over the
other (majority of them gather in the same state), thus decreasing
$\sigma_{t}$ on all time scales. To highlight this shift in Fig.~\ref{fig:stdAsymmScaling}
we have shown both scaling law used for the symmetric case (dashed
curve) and for the extremely asymmetric case (solid curve).

\begin{figure}
\begin{centering}
\includegraphics[width=0.7\textwidth]{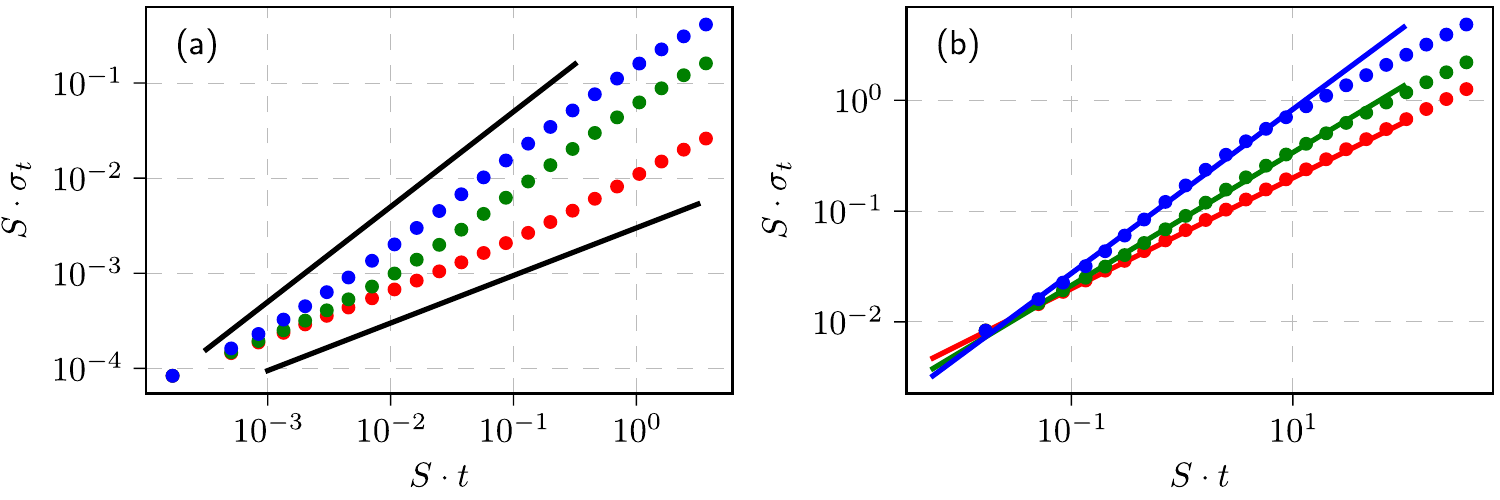}
\par\end{centering}
\caption{Scaling behavior of $\sigma_{t}$ in the model with impure intent
($q_{0}=1-q$ and $q_{1}=q$) in respect to $q$: numerical results
(colored dots) and power law fits (lines). Model parameter values:
$q=0.5$ (red dots), $0.57$ (green dots) and $0.68$ (blue dots),
$\varepsilon_{0}=\varepsilon_{1}=0.06$, $N=141$ (all in both cases),
$h=1.5\cdot10^{-4}$ (a) and $1.5\cdot10^{-2}$ (b). Power law fits
(a) have the following exponents: $\alpha=0.5$ (bottom curve) and
$1$ (upper curve). Power law fits (b): $\alpha=0.5$ (red curve),
$0.6$ (green curve) and $0.75$ (blue curve).\label{fig:stdFullScaling}}
\end{figure}

Finally let us also relax the pure intent assumption by assuming that
there is probability $q$ with which agent's intent is pure, i.e.,
let $q_{1}=q$ and $q_{0}=1-q$. Still this impure intent assumption
provides us a simplified version of the model as the impure intent
probability $q$ is assumed to be homogeneous. In general case, which
we will later use to fit the empirical data, $q_{0}$ and $q_{1}$
can take any value between $0$ and $1$ (as long as $q_{1}\geq q_{0}$).
The impure intent assumption is the final ingredient of the model,
which enables us to change the nature of diffusion in the short time
scale region. Though it is worth to note that, this is also the only
model mechanism which acts in discrete time. All other model mechanisms
would work the same way even if we would redefine the model in continuous
time, but it would impossible to redefine this mechanism for the continuous
time case without making any additional assumptions. Hence the impact
of $q$ is not trivial. For large values of $S$ it has no impact,
because the model is in the normal diffusion regime even with $q=1$.
For really small $S$ having $q<1$ introduces normal diffusion on
the shortest time scales, then on the intermediate time scales ballistic
regime is observed and finally on the longest time scales once again
normal diffusion takes over (see Fig.~\ref{fig:stdFullScaling} (a)).
For intermediate $S$ having $q<1$ allows for superdiffusive behavior
(see Fig.~\ref{fig:stdFullScaling} (b)), which we have observed
in the empirical data.

\begin{figure}
\begin{centering}
\includegraphics[width=0.4\textwidth]{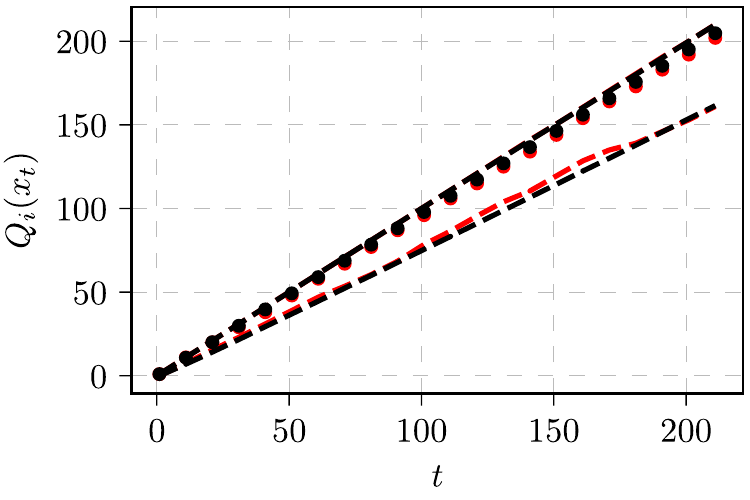}
\par\end{centering}
\caption{Presence quantile, $0.025$ (dashed curves), $0.5$ (points) and $0.975$
(dashed curves), series observed in the empirical data (red) and the
model (black). Empirical data averaged over both considered legislatures.
Model parameter values: $h=7.1\cdot10^{-4}$, $\varepsilon_{0}=0.21$,
$\varepsilon_{1}=0.43$, $q_{0}=0.8$, $q_{1}=0.98$.\label{fig:fullQuantiles}}
\end{figure}

\begin{figure}
\begin{centering}
\includegraphics[width=0.4\textwidth]{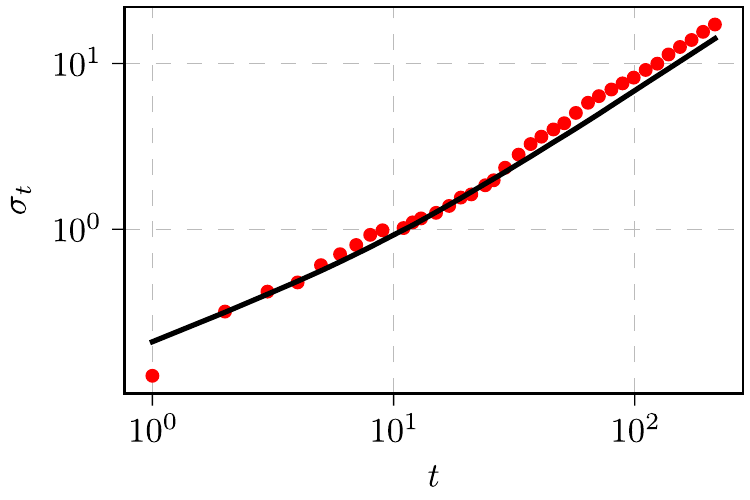}
\par\end{centering}
\caption{Standard deviation series $\sigma_{t}$ in the empirical data (red
dots) and the model (black curve). Empirical data averaged over both
considered legislatures. Model parameters are the same as in Fig.~\ref{fig:fullQuantiles}.\label{fig:fullStd}}
\end{figure}

\begin{figure}
\begin{centering}
\includegraphics[width=0.7\textwidth]{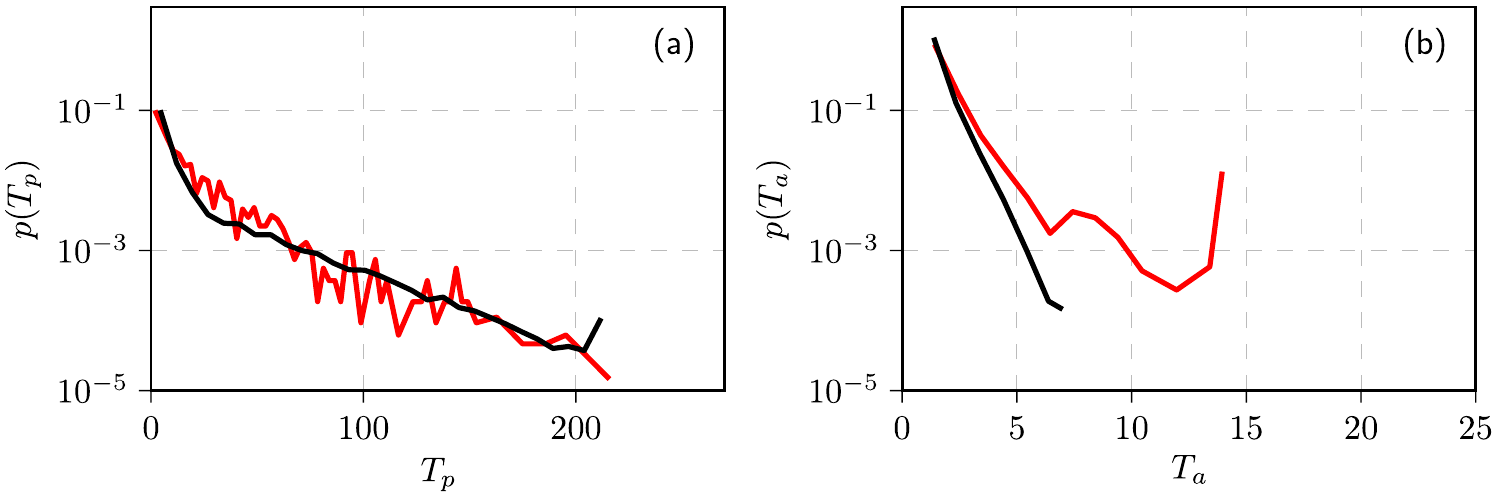}
\par\end{centering}
\caption{Probability density functions of the length of presence, $T_{p}$,
(a) and absence, $T_{a}$, (b) streaks in the empirical data (red
curves) and the model (black curves). Empirical data averaged over
both considered legislatures. Model parameters are the same as in
Fig.~\ref{fig:fullQuantiles}.\label{fig:fullStreaks}}
\end{figure}

This analysis provides some qualitative insights into the various
behaviors one could observe in the empirical presence data. Yet our
goal in this paper is to match not only $\sigma_{t}$ series, but
also presence quantile series (as a proxy for the temporal evolution
of the presence distribution) and attendance streak distributions.
Therefore we have performed random parameter space sweep, which was
somewhat informed by the previous analysis, and were able to find
model parameter set, which generates presence records with statistical
properties similar to those observed in the empirical data (see Figs.~\ref{fig:fullQuantiles},~\ref{fig:fullStd}~and~\ref{fig:fullStreaks}).
Presence quantiles, as expected, exhibit linear growth trends and
the overlap between the empirical data and the numerical results is
rather good (see Fig.~\ref{fig:fullQuantiles}). We were also able
to quite precisely reproduce the superdiffusive behavior of the records
as well (see Fig.~\ref{fig:fullStd}). Presence streak distribution
is also reproduced rather nicely (see Fig.~\ref{fig:fullStreaks}
(a)). The only evident disagreement between the model and the empirical
data is the absence streak distribution, which was observed to be
noticeably broader in the empirical data. This discrepancy might be
attributed to the small size of the empirical data, but also to a
more complicated dynamics of being absent. Namely, the proposed model
does not take into account specific circumstances, which are not directly
related to the social aspects of attendance. Such circumstances may
include sickness leaves, business or leisure trips, which in these
cases the representative would skip multiple sessions during that
period.

\section{Conclusions\label{sec:conclusions}}

We have analyzed the parliamentary presence data for the Lithuanian
parliament legislatures of 2008--2012 and 2012--2016. A similar
analysis was conducted earlier \cite{Vieira2019PRE} using Brazilian
parliamentary presence data. Unlike \cite{Vieira2019PRE} we haven't
found linear trend in standard deviation series (so--called ballistic
regime), but instead we have found sub--linear trend. Though the
trend is faster than would be expected from the normal diffusion,
therefore we can conclude that the considered empirical data exhibits
superdiffusive behavior. To complement the empirical analysis we have
also examined the distribution of the presence and absence streaks
in the data. We have found that both streak distributions are reasonably
close to an exponential distribution for the shortest time scales,
but we have also observed fatter tails.

To replicate selected statistical properties of the empirical data
we have built a simple agent--based model, which is built upon the
voter model. Unlike most models built upon the voter model, this model
involves two state dynamics, where one state is the true state (intent)
of the agent while the other is the observed state (action) of the
agent. While such voter models are not novel, the application and
the point of view taken here are. Voter models were not considered
from the anomalous diffusion point of view, because the whole system
trajectories evidently would not exhibit anomalous diffusion, but
here we have shown that distinct agent trajectories can. In our analysis
whenever we have considered only the true state dynamics, we have
observed only ballistic regime or normal diffusion. Adding the second,
noisy observation, state to the model have helped us to introduce
superdiffusion of the distinct agent trajectories. The proposed model
not only successfully reproduced the observed superdiffusive behavior,
but was also able to reproduce presence quantile series and attendance
streak distributions. The proposed model could also reproduce the
ballistic regime as observed in the Brazilian data set, though values
of $q_{0}$ and $h$ would likely be smaller (implying more truthfulness
as well as slower changing of the intent).

\section*{Acknowledgements}

Research was funded by European Social Fund (Project No 09.3.3-LMT-K-712-02-0026).


\begin{thebibliography}{10}
\expandafter\ifx\csname url\endcsname\relax
  \def\url#1{\texttt{#1}}\fi
\expandafter\ifx\csname urlprefix\endcsname\relax\def\urlprefix{URL }\fi
\expandafter\ifx\csname href\endcsname\relax
  \def\href#1#2{#2} \def\path#1{#1}\fi

\bibitem{Metzler2000PhysRep}
R.~Metzler, J.~Klafter, The random walk's guide to anomalous diffusion: a
  fractional dynamics approach, Physics Reports 339~(1) (2000) 1 -- 77.
\newblock \href {https://doi.org/10.1016/S0370-1573(00)00070-3}
  {\path{doi:10.1016/S0370-1573(00)00070-3}}.

\bibitem{Chepizko2013PRL}
O.~Chepizhko, F.~Peruani, Diffusion, subdiffusion, and trapping of active
  particles in heterogeneous media, Physical Review Letters 111 (2013) 160604.
\newblock \href {https://doi.org/10.1103/PhysRevLett.111.160604}
  {\path{doi:10.1103/PhysRevLett.111.160604}}.

\bibitem{Evers2013PRE}
F.~Evers, C.~Zunke, R.~D.~L. Hanes, J.~Bewerunge, I.~Ladadwa, A.~Heuer, S.~U.
  Egelhaaf, Particle dynamics in two-dimensional random-energy landscapes:
  {E}xperiments and simulations, Physical Review E 88 (2013) 022125.
\newblock \href {https://doi.org/10.1103/PhysRevE.88.022125}
  {\path{doi:10.1103/PhysRevE.88.022125}}.

\bibitem{Iaconis2019PRB}
J.~Iaconis, S.~Vijay, R.~Nandkishore, Anomalous subdiffusion from subsystem
  symmetries, Physical Review B 100 (2019) 214301.
\newblock \href {https://doi.org/10.1103/PhysRevB.100.214301}
  {\path{doi:10.1103/PhysRevB.100.214301}}.

\bibitem{Mercadier2009NatPhys}
N.~Mercardier, W.~Guerin, M.~Chevrollier, R.~Kaiser, L\'evy flights of photons
  in hot atomic vapours, Nature Physics 5 (2009) 602--605.
\newblock \href {https://doi.org/10.1038/nphys1286}
  {\path{doi:10.1038/nphys1286}}.

\bibitem{Scalliet2015PRL}
C.~Scalliet, A.~Gnoli, A.~Puglisi, A.~Vulpiani, Cages and anomalous diffusion
  in vibrated dense granular media, Physics Review Letters 114 (2015) 198001.
\newblock \href {https://doi.org/10.1103/PhysRevLett.114.198001}
  {\path{doi:10.1103/PhysRevLett.114.198001}}.

\bibitem{Kiselev2019PRL}
E.~I. Kiselev, J.~Schmalian, L\'evy flights and hydrodynamic superdiffusion on
  the {D}irac cone of graphene, Physical Review Letters 123 (2019) 195302.
\newblock \href {https://doi.org/10.1103/PhysRevLett.123.195302}
  {\path{doi:10.1103/PhysRevLett.123.195302}}.

\bibitem{Fogedby1994PRE}
H.~C. Fogedby, Langevin equations for continuous time {L\'evy} flights,
  Physical Review E 50 (1994) 1657--1660.
\newblock \href {https://doi.org/10.1103/PhysRevE.50.1657}
  {\path{doi:10.1103/PhysRevE.50.1657}}.

\bibitem{Baule2005PRE}
A.~Baule, R.~Friedrich, Joint probability distributions for a class of
  non-markovian processes, Physical Review E 71 (2005) 026101.
\newblock \href {https://doi.org/10.1103/PhysRevE.71.026101}
  {\path{doi:10.1103/PhysRevE.71.026101}}.

\bibitem{Kazakevicius2015PhysA}
R.~Kazakevicius, J.~Ruseckas, Anomalous diffusion in nonhomogeneous media:
  {P}ower spectral density of signals generated by time-subordinated nonlinear
  {L}angevin equations, Physica A 438 (2015) 210--222.
\newblock \href {https://doi.org/10.1016/j.physa.2015.06.047}
  {\path{doi:10.1016/j.physa.2015.06.047}}.

\bibitem{Ruseckas2016JStat}
J.~Ruseckas, R.~Kazakevicius, B.~Kaulakys, 1/f noise from point process and
  time-subordinated {L}angevin equations, Journal of Statistical Mechanics 2016
  (2016) 054022.
\newblock \href {https://doi.org/10.1088/1742-5468/2016/05/054022}
  {\path{doi:10.1088/1742-5468/2016/05/054022}}.

\bibitem{Cherstvy2013NJP}
A.~G. Cherstvy, A.~V. Chechkin, R.~Metzler, Anomalous diffusion and ergodicity
  breaking in heterogeneous diffusion processes, New Journal of Physics 15
  (2013) 083039.
\newblock \href {https://doi.org/10.1088/1367-2630/15/8/083039}
  {\path{doi:10.1088/1367-2630/15/8/083039}}.

\bibitem{Cherstvy2014SM}
A.~G. Cherstvy, A.~V. Chechkin, R.~Metzler, Particle invasion, survival, and
  non-ergodicity in {2D} diffusion processes with space-dependent diffusivity,
  Soft Matter 10 (2014) 1591--1601.
\newblock \href {https://doi.org/10.1039/C3SM52846D}
  {\path{doi:10.1039/C3SM52846D}}.

\bibitem{Kazakevicius2016PRE}
R.~Kazakevicius, J.~Ruseckas, Influence of external potentials on heterogeneous
  diffusion processes, Physical Review E 94 (2016) 032109.
\newblock \href {https://doi.org/10.1103/PhysRevE.94.032109}
  {\path{doi:10.1103/PhysRevE.94.032109}}.

\bibitem{Vieira2019PRE}
D.~S. Vieira, J.~M.~E. Riveros, M.~Jauregui, R.~S. Mendes, Anomalous diffusion
  behavior in parliamentary presence, Physical Review E 99 (2019) 042141.
\newblock \href {https://doi.org/10.1103/PhysRevE.99.042141}
  {\path{doi:10.1103/PhysRevE.99.042141}}.

\bibitem{Castellano2009RevModPhys}
C.~Castellano, S.~Fortunato, V.~Loreto, Statistical physics of social dynamics,
  Reviews of Modern Physics 81 (2009) 591--646.
\newblock \href {https://doi.org/10.1103/RevModPhys.81.591}
  {\path{doi:10.1103/RevModPhys.81.591}}.

\bibitem{Jedrzejewski2019CRP}
A.~Jedrzejewski, K.~Sznajd-Weron, Statistical physics of opinion formation:
  {I}s it a {SPOOF}?, Comptes Rendus Physique 20~(4) (2019) 244--261.
\newblock \href {https://doi.org/10.1016/j.crhy.2019.05.002}
  {\path{doi:10.1016/j.crhy.2019.05.002}}.

\bibitem{Redner2019CRP}
S.~Redner, Reality inspired voter models: a mini-review, Comptes Rendus
  Physique 20~(4) (2019) 275--292.
\newblock \href {https://doi.org/10.1016/j.crhy.2019.05.004}
  {\path{doi:10.1016/j.crhy.2019.05.004}}.

\bibitem{Mobilia2007JStatMech}
M.~Mobilia, A.~Petersen, S.~Redner, On the role of zealotry in the voter model,
  Journal of Statistical Mechanics: Theory and Experiment 2007~(08) (2007)
  P08029.
\newblock \href {https://doi.org/10.1088/1742-5468/2007/08/p08029}
  {\path{doi:10.1088/1742-5468/2007/08/p08029}}.

\bibitem{Khalil2018PRE}
N.~Khalil, M.~San~Miguel, R.~Toral, Zealots in the mean-field noisy voter
  model, Physical Review E 97 (2018) 012310.
\newblock \href {https://doi.org/10.1103/PhysRevE.97.012310}
  {\path{doi:10.1103/PhysRevE.97.012310}}.

\bibitem{Kirman1993QJE}
A.~P. Kirman, Ants, rationality and recruitment, Quarterly Journal of Economics
  108 (1993) 137--156.
\newblock \href {https://doi.org/10.2307/2118498} {\path{doi:10.2307/2118498}}.

\bibitem{Granovsky1995}
L.~B. Granovsky, N.~Madras, The noisy voter model, Stochastic Processes and
  their Applications 55~(1) (1995) 23--43.
\newblock \href {https://doi.org/10.1016/0304-4149(94)00035-R}
  {\path{doi:10.1016/0304-4149(94)00035-R}}.

\bibitem{Alfarano2009Dyncon}
S.~Alfarano, M.~Milakovic, Network structure and {N}-dependence in agent-based
  herding models, Journal of Economic Dynamics and Control 33~(1) (2009)
  78--92.
\newblock \href {https://doi.org/10.1016/j.jedc.2008.05.003}
  {\path{doi:10.1016/j.jedc.2008.05.003}}.

\bibitem{Kononovicius2014EPJB}
A.~Kononovicius, J.~Ruseckas, Continuous transition from the extensive to the
  non-extensive statistics in an agent-based herding model, European Physics
  Journal B 87~(8) (2014) 169.
\newblock \href {https://doi.org/10.1140/epjb/e2014-50349-0}
  {\path{doi:10.1140/epjb/e2014-50349-0}}.

\bibitem{Carro2016}
A.~Carro, R.~Toral, M.~San~Miguel, The noisy voter model on complex networks,
  Scientific Reports 6 (2016) 24775.
\newblock \href {https://doi.org/10.1038/srep24775}
  {\path{doi:10.1038/srep24775}}.

\bibitem{Peralta2018Chaos}
A.~F. Peralta, A.~Carro, M.~San~Miguel, R.~Toral, Analytical and numerical
  study of the non-linear noisy voter model on complex networks, Chaos 28
  (2018) 075516.
\newblock \href {https://doi.org/10.1063/1.5030112}
  {\path{doi:10.1063/1.5030112}}.

\bibitem{Mori2019PRE}
S.~Mori, M.~Hisakado, K.~Nakayama, Voter model on networks and the multivariate
  beta distribution, Physical Review E 99 (2019) 052307.
\newblock \href {https://doi.org/10.1103/PhysRevE.99.052307}
  {\path{doi:10.1103/PhysRevE.99.052307}}.

\bibitem{Gastner2019JPA}
M.~T. Gastner, K.~Ishida, Voter model on networks partitioned into two cliques
  of arbitrary sizes, Journal of Physics A 52 (2019) 505701.
\newblock \href {https://doi.org/10.1088/1751-8121/ab542f}
  {\path{doi:10.1088/1751-8121/ab542f}}.

\bibitem{Masuda2010PRE}
N.~Masuda, N.~Gibert, S.~Redner, Heterogeneous voter models, Physical Review E
  82 (2010) 010103.
\newblock \href {https://doi.org/10.1103/PhysRevE.82.010103}
  {\path{doi:10.1103/PhysRevE.82.010103}}.

\bibitem{Gastner2018JStat}
M.~T. Gastner, B.~Oborny, M.~Glyas, Consensus time in a voter model with
  concealed and publicly expressed opinions, Journal of Statistical Mechanics
  2018 (2018) 063401.
\newblock \href {https://doi.org/10.1088/1742-5468/aac14a}
  {\path{doi:10.1088/1742-5468/aac14a}}.

\bibitem{Jedrzejewski2018PlosOne}
A.~Jedrzejewski, G.~Marcjasz, P.~R. Nail, K.~Sznajd-Weron, Think then act or
  act then think?, PLOS ONE 13~(11) (2018) 1--19.
\newblock \href {https://doi.org/10.1371/journal.pone.0206166}
  {\path{doi:10.1371/journal.pone.0206166}}.

\bibitem{Artime2019CRPhys}
O.~Artime, A.~Carro, A.~F. Peralta, J.~J. Ramasco, M.~San~Miguel, R.~Toral,
  Herding and idiosyncratic choices: nonlinearity and aging-induced transitions
  in the noisy voter model, Comptes Rendus Physique (2019).
\newblock \href {https://doi.org/10.1016/j.crhy.2019.05.003}
  {\path{doi:10.1016/j.crhy.2019.05.003}}.

\bibitem{Castellano2009PRE}
C.~Castellano, M.~A. Munoz, R.~Pastor-Satorras, The non-linear q-voter model,
  Physical Review E 80 (2009) 041129.
\newblock \href {https://doi.org/10.1103/PhysRevE.80.041129}
  {\path{doi:10.1103/PhysRevE.80.041129}}.

\bibitem{FernandezGarcia2014PRL}
J.~Fernandez-Gracia, K.~Suchecki, J.~J. Ramasco, M.~San~Miguel, V.~M. Eguiluz,
  Is the voter model a model for voters?, Physical Review Letters 112 (2014)
  158701.
\newblock \href {https://doi.org/10.1103/PhysRevLett.112.158701}
  {\path{doi:10.1103/PhysRevLett.112.158701}}.

\bibitem{Sano2016}
F.~Sano, M.~Hisakado, S.~Mori, Mean field voter model of election to the house
  of representatives in {J}apan, in: JPS Conference Proceedings, Vol.~16, The
  Physical Society of Japan, 2017, p. 011016.
\newblock \href {https://doi.org/10.7566/JPSCP.16.011016}
  {\path{doi:10.7566/JPSCP.16.011016}}.

\bibitem{Kononovicius2017Complexity}
A.~Kononovicius, Empirical analysis and agent-based modeling of {L}ithuanian
  parliamentary elections, Complexity 2017 (2017) 7354642.
\newblock \href {https://doi.org/10.1155/2017/7354642}
  {\path{doi:10.1155/2017/7354642}}.

\bibitem{Braha2017PlosOne}
D.~Braha, M.~A.~M. de~Aguiar, Voting contagion: Modeling and analysis of a
  century of u.s. presidential elections, PLOS ONE 12~(5) (2017) 1--30.
\newblock \href {https://doi.org/10.1371/journal.pone.0177970}
  {\path{doi:10.1371/journal.pone.0177970}}.

\bibitem{Kononovicius2019CompJStat}
A.~Kononovicius, Compartmental voter model, Journal of Statistical Mechanics
  2019 (2019) 103402.
\newblock \href {https://doi.org/10.1088/1742-5468/ab409b}
  {\path{doi:10.1088/1742-5468/ab409b}}.

\bibitem{Alfarano2005CompEco}
S.~Alfarano, T.~Lux, F.~Wagner, Estimation of agent-based models: {T}he case of
  an asymmetric herding model, Computational Economics 26~(1) (2005) 19--49.
\newblock \href {https://doi.org/10.1007/s10614-005-6415-1}
  {\path{doi:10.1007/s10614-005-6415-1}}.

\bibitem{Alfarano2008Dyncon}
S.~Alfarano, T.~Lux, F.~Wagner, Time variation of higher moments in a financial
  market with heterogeneous agents: {A}n analytical approach, Journal of
  Economic Dynamics and Control 32 (2008) 101--136.
\newblock \href {https://doi.org/10.1016/j.jedc.2006.12.014}
  {\path{doi:10.1016/j.jedc.2006.12.014}}.

\bibitem{Kononovicius2012PhysA}
A.~Kononovicius, V.~Gontis, Agent based reasoning for the non-linear stochastic
  models of long-range memory, Physica A 391~(4) (2012) 1309--1314.
\newblock \href {https://doi.org/10.1016/j.physa.2011.08.061}
  {\path{doi:10.1016/j.physa.2011.08.061}}.

\bibitem{Gontis2014PlosOne}
V.~Gontis, A.~Kononovicius, Consentaneous agent-based and stochastic model of
  the financial markets, PLoS ONE 9~(7) (2014) e102201.
\newblock \href {https://doi.org/10.1371/journal.pone.0102201}
  {\path{doi:10.1371/journal.pone.0102201}}.

\bibitem{Franke2018}
R.~Franke, F.~Westerhoff, Different compositions of animal spirits and their
  impact on macroeconomic stability, working Paper at University of Bamberg.
\newblock \href {https://doi.org/10.13140/RG.2.2.26068.09609}
  {\path{doi:10.13140/RG.2.2.26068.09609}}.

\bibitem{Kononovicius2019OB}
A.~Kononovicius, J.~Ruseckas, Order book model with herding behavior exhibiting
  long-range memory, {Physica A} 525 (2019) 171--191.
\newblock \href {https://doi.org/10.1016/j.physa.2019.03.059}
  {\path{doi:10.1016/j.physa.2019.03.059}}.

\bibitem{Vilela2019PhysA}
A.~L.~M. Vilela, C.~Wang, K.~P. Nelson, H.~E. Stanley, Majority-vote model for
  financial markets, Physics A 515 (2019) 762--770.
\newblock \href {https://doi.org/10.1016/j.physa.2018.10.007}
  {\path{doi:10.1016/j.physa.2018.10.007}}.

\bibitem{LRSData}
{Lietuvos Respublikos Seimas},
  \href{https://www.lrs.lt/sip/portal.show?p_r=35391&p_k=1}{Atviri duomenys},
  accessed on 2019-06-08.
\newline\urlprefix\url{https://www.lrs.lt/sip/portal.show?p_r=35391&p_k=1}

\bibitem{Kononovicius2020GitPresence}
A.~Kononovicius,
  \href{https://github.com/akononovicius/lithuanian-parliamentary-presence-data}{Lithuanian
  parliamentary presence data}, Github repository (2020).
\newline\urlprefix\url{https://github.com/akononovicius/lithuanian-parliamentary-presence-data}

\bibitem{Clifford1973}
P.~Clifford, A.~Sudbury, A model for spatial conflict, Biometrika 60 (1973) 581
  -- 588.
\newblock \href {https://doi.org/10.1093/biomet/60.3.581}
  {\path{doi:10.1093/biomet/60.3.581}}.

\bibitem{Liggett1999}
T.~Liggett, Stochastic interacting systems: {C}ontact, voter, and exclusion
  processes, Springer, 1999.

\bibitem{Willis1965}
H.~R. Willis, Conformity, independence and anticonformity, Human Relations 18
  (1965) 373.
\newblock \href {https://doi.org/10.1177/001872676501800406}
  {\path{doi:10.1177/001872676501800406}}.

\bibitem{Nail2016APPA}
P.~R. Nail, K.~Sznajd-Weron, The diamond model of social response within an
  agent-based approach, Acta Physica Polonica A 129~(5) (2016) 1050--1054.
\newblock \href {https://doi.org/10.12693/APhysPolA.129.1050}
  {\path{doi:10.12693/APhysPolA.129.1050}}.

\bibitem{Rabiner1989IEEE}
L.~R. {Rabiner}, A tutorial on hidden {M}arkov models and selected applications
  in speech recognition, Proceedings of the IEEE 77~(2) (1989) 257--286.
\newblock \href {https://doi.org/10.1109/5.18626} {\path{doi:10.1109/5.18626}}.

\bibitem{Gillespie1977JPC}
D.~T. Gillespie, Exact stochastic simulation of coupled chemical reactions,
  Journal of Physical Chemistry 81 (1977) 2340--2361.
\newblock \href {https://doi.org/10.1021/j100540a008}
  {\path{doi:10.1021/j100540a008}}.

\bibitem{Carro2015PlosOne}
A.~Carro, R.~Toral, M.~San~Miguel, Markets, herding and response to external
  information, PLoS ONE 10 (2015) e0133287.
\newblock \href {https://doi.org/10.1371/journal.pone.0133287}
  {\path{doi:10.1371/journal.pone.0133287}}.

\bibitem{Goldenberg2001}
J.~Goldenberg, B.~Libai, E.~Muller, Using complex systems analysis to advance
  marketing theory development, Academy of Marketing Science Review 9 (2001)
  1--18.

\bibitem{Daniunas2011ICCGI}
V.~Daniunas, V.~Gontis, A.~Kononovicius, Agent-based versus macroscopic
  modeling of competition and business processes in economics, in: ICCGI 2011,
  The Sixth International Multi-Conference on Computing in the Global
  Information Technology, Luxembourg, 2011, pp. 84--88.

\bibitem{Martins2008IJMPC}
A.~C.~R. Martins, Continuous opinions and discrete actions in opinion dynamics
  problem, International Journal of Modern Physics C 19~(4) (2008) 617--624.
\newblock \href {https://doi.org/10.1142/S0129183108012339}
  {\path{doi:10.1142/S0129183108012339}}.

\bibitem{Kononovicius2020GitVM}
A.~Kononovicius,
  \href{https://github.com/akononovicius/noisy-voter-model-for-parliamentary-presence}{Noisy
  voter model for the parliamentary presence}, Github repository (2020).
\newline\urlprefix\url{https://github.com/akononovicius/noisy-voter-model-for-parliamentary-presence}

\end{thebibliography}
\end{document}